# Decoding the Workplace & EOR: An Employee Survey Analysis by Data Science Techniques and Visualization

Kishankumar Bhimani [*] [0000-0002-9792-1484] and Khushbu Saradva [0000-0003-2732-3478]

Laboratory on AI for Computational Biology, Department of Data Analysis and Artificial Intelligence, Faculty of Computer Science, HSE University, 11 Pokrovsky Boulevard, 109028, Moscow, Russian Federation
`kbhimani@hse.ru`

**Abstract.** This research study explores the new dynamics of employee-organization relationships (EOR) [6] using advanced data science methodologies and presents findings through accessible visualizations. Leveraging a dataset procured from a comprehensive nationwide big employee survey, this study employs innovative strategy for theoretical researcher by using our state-of-the-art visualization. The results present insightful visualizations encapsulating demographic analysis, workforce satisfaction, work environment scrutiny, and the employee's view via word cloud interpretations and burnout predictions.

The study underscores the profound implications of data science across various management sectors, enhancing understanding of workplace dynamics and promoting mutual growth and satisfaction. This multifaceted approach caters to a diverse array of readers, from researchers in sociology and management to firms seeking detailed understanding of their workforce's satisfaction, emphasizing on practicality and interpretability.

The research encourages proactive measures to improve workplace environments, boost employee satisfaction, and foster healthier, more productive organizations. It serves as a resourceful tool for those committed to these objectives, manifesting the transformative potential of data science in driving insightful narratives about workplace dynamics and employee-organization relationships. In essence, this research unearths valuable insights to aid management, HR professionals, and companies in harnessing data science techniques for their future research endeavors.

## 1      Introduction

The comprehension of workplace dynamics [21] is an area of study that has gathered significant attention, given its intricate connection with organizational success. This comprehension, which involves understanding the complexities of employee behaviors and their linkage with organizational elements, is critical to fostering a conducive and productive work environment.

This paper delves into a comprehensive analysis of the workplace, employing data science methodologies and visualization techniques to a dataset gleaned from a large employee survey. We have undertaken this study with a keen interest in understanding the multifaceted relationship between employees and their organization. We focus on several key facets including employee satisfaction, their engagement levels, work environment conditions, demographic analysis, and more critically, burnout prediction. This multifocal approach enables a holistic understanding of the workplace, thereby furnishing actionable insights.

Additionally, this research aims to bridge the gap between complex data science methods and their practical application in understanding employee-organization dynamics. We present our analysis in an accessible manner, thereby enabling individuals even with limited programming or data science knowledge to comprehend the insights derived. One can get benefit by using such approach who work in EOR.

Through this paper, we strive to contribute to the broader dialogue on employee-organization relationships and the various elements influencing it. We are hopeful that our research will serve as a valuable key for various stakeholders, including HR professionals, organizational leaders, researchers, and employees alike. We believe that the insights derived from our analysis can catalyze effective decision-making processes and lead to the development of healthier, more productive workplaces.

## 2    Literature Review

For decades, scholars specializing in organizational behavior have dedicated themselves to a deeper understanding of the relationship between employees and their respective organizations [12]. The Employee-Organization Relationship (EOR) has steadily emerged as a central area of study for experts in organizational behavior, human resource management, and industrial relations. Existing literature comprehensively explores the EOR from both the individual perspective and at group and organizational levels of analysis.

The EOR has entrenched itself in the academic discourse, seeking to underpin the theoretical groundwork essential for comprehending the mutual exchange from both employer and employee standpoints. The relationship between employee performance and employer benefits represents a multifaceted dynamic [2]. There exists a mutual commitment to improvement on both ends of this spectrum, which can be stimulated through various forms of analysis. Numerous methodologies, approaches, and publications provide substantial literature in this domain. They address key areas such as the employee-organization relationship, employee satisfaction, burnout, workspace environment, and the consequential link between employee engagement and organizational performance.

The paper by Frye et al. [9] underscores the significant financial and operational costs associated with employee attrition. Their research highlights the importance of predictive models, such as logistic regression, in understanding and preventing employee quits. This study aligns with a growing body of literature emphasizing the need to address employee attrition and burnout for organizational success. Brewer and Shapard's [4] study conducts a meta-analysis on the connection between employee burnout and factors like age and experience. It underscores the practical significance of understanding this relationship for effective burnout interventions. The literature review emphasizes the harmful impact of burnout on individuals and organizations and the need to reconcile differing research findings to guide HR professionals in addressing burnout effectively.

Nair et al. [16] paper underscores the indispensable role of data visualization in predictive analytics, highlighting how visual analytics, in tandem with various Data Science tools like, bridging the gap between raw data and actionable insights of business intelligence and this is the key to expand this to real employee survey visual analytics.

This body of literature acts as a comprehensive resource for understanding the nuances and dynamics of the employee-organization relationship, facilitating the formulation of effective strategies for mutual growth and satisfaction. Here, Data science could lead this field by analysis and visualization [19] because it covers wide range of solution for all sectors. The contemporary workplace is undergoing significant transformations driven by technological advancements, changing demographics, and evolving employee expectations. Organizations are increasingly recognizing the importance of Employee Opinion and Engagement (EOE) to ensure their success and sustainability. In this context, the application of data science techniques and visualization tools to analyze employee survey data presents a valuable opportunity for organizations to gain insights into employee perceptions, sentiments, and engagement levels. This literature review aims to explore existing research on the intersection of workplace analysis, Employee Opinion and Engagement (EOE), data science methodologies, and visualization techniques.

## 3    Data

This paper aims to elaborate on the dynamics of today's workplaces through the application of advanced data science techniques. The backbone of this extensive analysis rests on a meticulously collected, broad-ranging data set procured from a nationwide employee survey.

This facts set, sourced through a paid acquisition rather than open-source means, gives an inclusive view of the contemporary place of wide landscape, encompassing all industry sectors and employee roles. The data size is substantial, originally composed of 18,976 unique entries with 82 attributes each. However, through a process of cleaning and removal of duplicate entries, this raw dataset was streamlined to 18,972 unique records with 80 survey features, further enhancing the potential for precise insights. Since the information predominantly consists of survey responses, it necessitated rigorous information wrangling and filtering tactics to streamline it for significant interpretation and analysis [20].

To ensure transparency and to support reproducibility in future research, we will facilitate access to a sample copy of our dataset in a secure manner. It protects the privacy of research participants and aims to help other researchers validate our methods and results while adhering to data protection standards. The aim of sharing this sample data is to foster a collaborative, transparent research environment that would encourage further explorations in this domain.

### 3.1     Data Collection

The data used in this paper is accurately acquired from employee survey based firm from specific country. The motivation behind this collection strategy was to build a comprehensive model to elucidate workplace dynamics for both employees and organizations. To ensure the reliability and accuracy of our data, we opted to source this large dataset from a reputable organization specializing in high-quality, precision data surveys.

The structure of the data collection is described in detail in **Table 1**. The table represents the raw data from the employee survey, where the headers define the questions tag, and the responses are recorded in subsequent rows of the dataframe. This layout provides a clear and systematic view of the data for an effective understanding to learners.

### 3.2     Data Pre-processing

Preprocessing the collected data is a key to ensuring its readiness for analysis [15]. The following methods were systematically used for this purpose.

- The initial data headers contained redundant tags i.e. <br> that could complicate analysis. We rectified this by transforming these tags into simplified dataframe columns, thereby enhancing the clarity of the dataset.
- Handling missing value and data steam line processing for ready to use.
- Responses to several survey questions were taken on a five-point scale: *'Agree', 'Rather agree than disagree', 'Disagree rather than agree', 'Disagree', and 'Difficult to answer'*. We numericalized these responses, mapping them to the values 1, 2, 3, 4, and 5, respectively for bidirectional conversion.
- We applied label encoding, scaling, and various transformations during the programming phase to ensure standardization and enhance the reliability of the dataset.
- A range of data science techniques were applied to further prepare the dataframe. These included methods for data mapping, transformation, and others, all designed to enhance the usability of the dataset.
- In preparation for text processing tasks such as word cloud creation and burn prediction, we performed stop word removal using the NLTK library. Stop words, which are commonly occurring words of minimal significance (e.g., 'a', 'an', 'the', 'are'), were removed to focus the text analysis on more meaningful content.

Through these comprehensive pre-processing steps, we were able to refine our dataset, increasing its suitability for a thorough and accurate workplace analysis.

**Table 1.** Employee Survey Data with in detailed features, questions and responses

| Sr. No. | Question Tag | Response |
|---|---|---|
| 1 | Key (value) | Unique value |
| 2 | Session ID | Unique Session Data |
| 3 | Company | Individual Company name |
| 4 | Industry | Industry Type |
| 5 | Number of staff | Range |
| 6 | start date | Survey started on |
| 7 | Date of completion | Survey completed on |
| 8 | Fill time (Years: Months: Days: Hours: Minutes: Seconds) | Total Time taken |
| 9 | With what probability from 0 to 10 are you ready to recommend your current employer as a place of work to your friends and acquaintances? | 1,2,3,4,5,6,7,8,9,10 1 - Lowest, 10 - Highest |
| 10 | I know the mission and strategy of our Company. | |
| 11 | I understand how the results of my work affect the overall result of the Company's work. | |
| 12 | I do not lack the information necessary for my work. | |
| 13 | In our Company, the work is organized efficiently and optimally regulated. | |
| 14 | My area of responsibility and tasks are clearly defined. | |
| 15 | The company provides me with all the necessary resources to do my job effectively. | |
| 16 | My work benefits people. | |
| 17 | I am interested in the work that I do in the Company. | |
| 18 | My work allows me to maximize my abilities. | |
| 19 | When I do my job, I feel energized. | Agree |
| 20 | I feel burned out, constantly tired and irritated from work. | Rather agree than disagree |
| 21 | Partnership constructive relations have been established between various divisions of our Company. | Disagree rather than agree |
| 22 | I have the opportunity to convey any information, including negative, to management. | Disagree |
| 23 | I know what I need to do to advance my career. | Difficult to answer |
| 24 | In our Company, initiative and efficient employees have the opportunity for career growth. | |
| 25 | I can earn more if I work more intensively and efficiently. | |
| 26 | The level of my income corresponds to my professional level and personal contribution to the overall result. | |
| 27 | I understand how our Company evaluates the effectiveness of my work. | |
| 28 | I am satisfied with the social package that the Company offers to its employees. | |
| 29 | Our Company provides comfortable working conditions (salary and social package) for its employees. | |
| 30 | I get enough recognition and praise when I do my job well. | |
| 31 | Our Company motivates employees to achieve maximum results. | |
| 32 | Our Company values and retains highly professional employees. | |
| 33 | The values of our Company are aligned with my personal values. | |
| 34 | In my daily work, I adhere to the values of the Company. | |
| 35 | I can achieve the elimination of barriers in the organization of work. | |
| 36 | Our Company encourages initiatives aimed at improving work. | |
| 37 | I can independently make decisions within my area of responsibility without excessive control from my manager. | |
| 38 | I freely, and without fear of negative consequences, can express my opinion. | |
| 39 | In my work, I try to do more than what is required of me. | |
| 40 | I like the atmosphere in our Company. | |

| Sr. No. | Question Tag | Response |
|---|---|---|
| 41 | I trust the professionalism of my colleagues and turn to them for advice in complex matters. | |
| 42 | In our Company, there are friendly and positive relations between employees. | |
| 43 | My immediate supervisor successfully organizes the work of our unit. | |
| 44 | My supervisor is the informal leader in our team. | |
| 45 | My manager supports the suggestions of employees to optimize work and helps to implement them. | |
| 46 | I trust the decisions made by the top managers of our Company. | |
| 47 | Our Company always fulfils its obligations to employees. | |
| 48 | I recommend our Company to my friends and acquaintances as a good employer. | |
| 49 | I plan to leave the Company within the next year. | |
| 50 | When asked, I proudly name the Company I work for. | Agree |
| 51 | I am ready to take on tasks that are unpleasant for me if necessary for the success of the Company. | Rather agree than disagree |
| 52 | I care about the successes and difficulties of my Company, even if they do not affect me personally. | Disagree rather than agree |
| 53 | I am ready to make significant efforts to make a career in the Company. | Disagree |
| 54 | I am ready to take responsibility for solving non-standard work situations. | Difficult to answer |
| 55 | In unusual working situations, I prefer not to take responsibility for the decision myself. | |
| 56 | I try not to take on tasks that are outside my area of responsibility. | |
| 57 | I try to find more effective ways to solve my work tasks. | |
| 58 | My direct supervisor treats subordinates with respect. | |
| 59 | My line manager regularly provides me with feedback on the results of my work. | |
| 60 | In our team, there are friendly and positive relations between employees. | |
| 61 | Working in the company gives me a sense of stability and confidence in the future. | |
| 62 | I get enough information about what is happening in our Company. | |
| 63 | Friendly and positive relations have been established between the various divisions of our Company. | |
| 64 | Efficient interaction has been established between divisions in our Company. | |
| 65 | Top management clearly explains the Company's strategy and goals and how to achieve them. | |
| 66 | I feel that the company cares about my health. | |
| 67 | I commend the actions taken by the management of our company in the context of the COVID-19 pandemic. | |
| 68 | What is the most important thing for you in your work? | Categories selection |
| 69 | Do you think that next year the position of your Company will become better or worse? | Better/ Worse comparison |
| 70 | Do you think your family will live better or worse next year? | Text response by employee |
| 71 | What do you think makes our company a good employer? | Text response by employee |
| 72 | What do you think should be improved in the work/processes in the company? | Yes/ No |
| 73 | Do you work remotely? | Position level |
| 74 | Your management level: | Experience Range |
| 75 | Your work experience in the Company: | Male / Female |
| 76 | What's your gender: | Age Range |
| 77 | Your age: | Region / Province of country |
| 78 | Where do you work? | Production category |
| 79 | Choose your function in the Company | Education range level |
| 80 | Please indicate your level of education | |

## 4    Methodology

This Methodology seeks to address an innovative approach to understand and navigate the complexities of employee surveys and workplace analytics, a challenge faced by numerous organizations in their quest for customized insights. We have given special consideration to visual interpretation by applying modern data science methodologies, facilitating an intuitive understanding of the analyses presented.

Our approach is designed to be accessible not only to those experienced in programming and data science but also to individuals who might be beginner in these domains. We provide ready-to-use codes, promoting wider adoption of the methods introduced.

The insights gained from the employee-organization dynamic are of vital importance to a range of disciplines, including Management, Economics, Sociology, and Human Resource. These sectors can leverage our visually rich approach to glean new insights, potentially enhancing their understanding of their domains and enriching future publications. In this section, we will elucidate our methodology, which comprises three main stages: Exploratory Data Analysis, Feature Engineering, and Predictive Analysis.

### 4.1    Exploratory Data Analysis

Any survey-based research must begin with exploratory data analysis (EDA) [14], which offers an excellent way to fully understand data patterns and linkages. By engaging in EDA, we can conduct statistical analyses, identify correlations, spot missing data, detect outliers, and determine the central tendency of the dataframe. These steps collectively permit us to find insightful patterns and informative visualizations.

EDA graphs, in their simplicity and expressiveness, provide a practical and effective means to represent the underlying data. EDA equips the researcher to discover the data's inherent structure by uncovering new revealing insights. In this study, the broad objectives of EDA are to enhance our understanding of the survey responses, identify patterns within the data and build easy to understand for audience.

Applied to our context, EDA plays a pivotal role in insights about employee satisfaction and engagement, organizational work culture, happiness indices, benefits offered, the company's value to its employees, and a demographic factors. The use of EDA in our methodology ensures that we capitalize on the richness of our dataset, uncovering nuanced information to inform our further analyses and discussions.

*Correlation analysis:* Correlation analysis is a vital aspect of (EDA) for understanding relationships between variables and quantifying their strength. In **Fig. 1**, we present a correlation analysis of 60 key questions from **Table 1**, focusing on response-based categorical data. The figure consists of subplots labeled from (A) to (F).

In (A) *Agree* Responses and (D) *Disagree* Responses, we observe normal distributions, indicating a balanced relationship between these categories. However, (B) *Rather Agree than Disagree* and (E) *Disagree* show skewed distributions, suggesting an asymmetry in the relationships. (C) *Difficult to Answer* displays a symmetric distribution, indicating a different pattern altogether. Subplot (F) is a correlation histogram, filled with p-values. Notably, this histogram displays aligned foundation to further analysis.

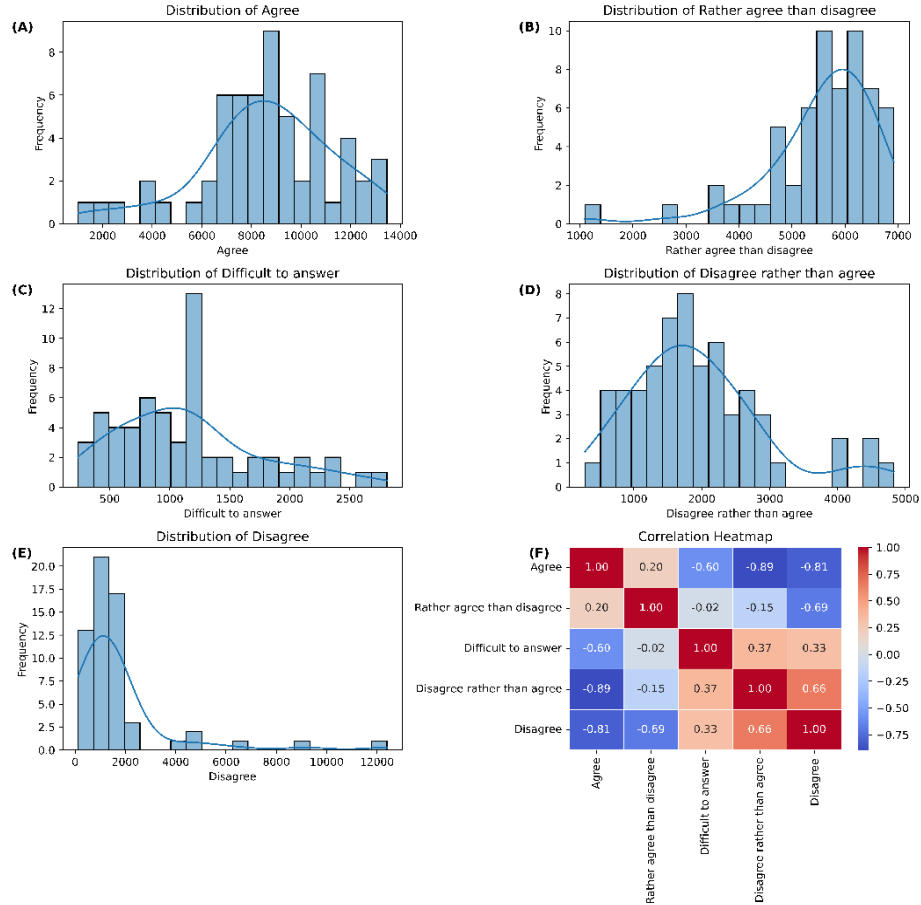

**Fig. 1** (A-E) Correlation analysis of response based categorical data. (F) Heatmap of correlation with p-value

### 4.2 Feature Engineering

In many instances, datasets, including those derived from questionnaires, articles, reviews, and surveys, contain textual fields that can be rich sources of information. Feature engineering is the process through which these raw data elements are carefully selected, manipulated, and transformed into more meaningful attributes or "features". These features, which can be quantified inputs in a predictive model, often significantly enhance the quality and accuracy of the resulting analysis [3].

A key strength of feature engineering is its role as a springboard to predictive analysis, enabling the application of Machine Learning (ML), Deep Learning (DL), and Natural Language Processing (NLP) techniques. From **Table 1**, questions 71 and 72 allow employees to express their perceptions freely. NLP techniques are pivotal in understanding individual sentiments. In **Fig.7**, we illustrate a commonly used Word Cloud approach. Researchers can further enrich this categorical data by employing NLP algorithms for text classification, such as Word2Vec, Glove, TF-IDF, Bag-of-Words, and Embedding Algorithms [11]. It can provide a predictive perspective, facilitating the forecasting and estimating of future outcomes.

The journey of feature engineering starts with data preprocessing and cleaning. We employ standard techniques such as data imputation for missing values using algorithms like mean imputation, median imputation, or advanced techniques like K-nearest neighbors (KNN) imputation [1]. Very few Outliers are identified and addressed using methods such as IQR-based detection [24]. One of the best use case we showed of feature engineering in **Fig.8**, here we used conditional based approach to select features.

Feature engineering can aid in interpreting survey trends and directing predictive analyses in the context of employee surveys. For instance, the replies "*Agree*", "*Rather agree than disagree*", "*Disagree rather than agree*", "*Difficult to answer*", and "*Agree*" on the five-point scale were converted into numerical features which was converted using mapping [7]. Now, we have numerical values representing each response, we can create features based on the survey questions. For example, let's say we have a survey question like, "I am satisfied with my job." We can create a feature called "Job_Satisfaction" and assign it the numerical values corresponding to the responses of each employee. These features, mapped against variables such as age, gender, industry, position, and region of work, allowed us to explore and identify patterns more effectively.

### 4.3   Predictive Analytics

Predictive analytics encompasses a range of business intelligence technologies designed to identify patterns and relationships within large datasets, enabling the prediction of future behaviors and events. In the context of our study, predictive analytics offers significant value by forecasting potential employee attrition. This insight can guide organizations to enhance their employee satisfaction, happiness, and benefits programs proactively, mitigating potential turnover [8].

Ponnuru et al. [18] present Machine Learning algorithms, specifically Logistic Regression, on IBM HR data to predict and address employee attrition for only software industry. Najafi [17] uses a machine learning framework with a "max-out" feature selection method to predict employee attrition and assess model stability using an IBM HR dataset, Srivastava et al. [22] utilizes predictive analytics to forecast and mitigate employee attrition by analyzing past data to predict future voluntary terminations.

In the realm of predictive analytics, we've established three vital components to gauge attrition score calculation as shown in **Fig.8**. First, we measure Employee Satisfaction (depicted in blue) by tallying the "Agree" responses in the Industry sector, directly extracted from Que_tag No 28. Second, we pinpoint Employee Burnout (illustrated in red) by identifying the column with the highest count of "Disagree" responses, Que_tag No 49, and calculating the "Agree" responses within it, which indicates employees considering leaving within the next year. Third, our custom-developed Employee Attrition score, created through condition-based programming, assesses the risk of employees leaving. This score takes into account responses from both Questions, capturing instances where employees express a desire to leave alongside dissatisfaction.

These components collectively provide a comprehensive understanding of potential burnout and attrition risks within the organization.

## 5      Results

### 5.1      Demographic Analysis & Visualization

In this findings, the demographic analysis plays a significant role, given the multiple features available for exploration and visualization within the dataset.

**Fig. 2** presents a pie chart representing the number of employees across different industry sectors. Each sector's percentage and total employee count are clearly displayed, providing a comprehensive view of employee distribution across these industry segments. The legend lists cover all industry sectors included in the dataset.

The primary objective of this visualization is to provide a clear and concise overview of how employees are distributed across various industry sectors. A Pie chart is particularly well-suited for this purpose as it allows for a quick and intuitive comparison of the percentage count of employees working within each industry sector relative to the total workforce. A Pie chart in this context due to its inherent strength in displaying the proportional composition of parts within a whole. Consequently, it facilitates a rapid assessment of the industry sectors with the highest and lowest proportions of employees, a key aspect of our analysis in understanding the workforce landscape.

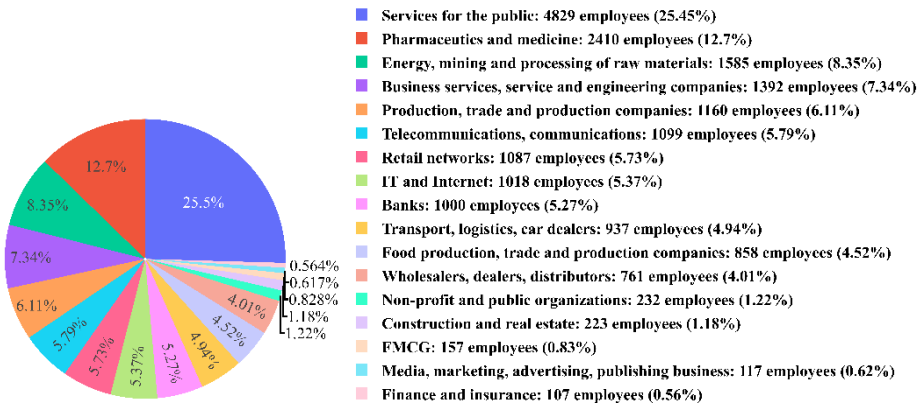

**Fig. 2.** A Pie chart visualizing distribution of employees across various Industry sectors

**Fig.3** demonstrates a bar chart illustrating the distribution of employees across various age ranges. The X-axis lists age group categories—ranging from "Up to 20 years old" to "Over 60 years old"—while the Y-axis reflects the corresponding employee counts. This visualization sheds valuable insights into the age demographics of the workforce, shedding light on the proportion of younger versus older employees across sectors.

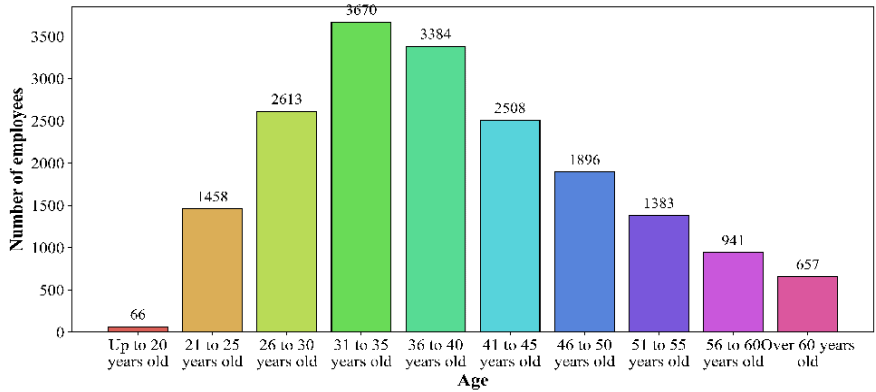

**Fig.3.** Employee age range distribution with total count

The objective of this visualization is to provide an overview of how the employee population is distributed across various age groups. A Bar chart is particularly well-suited for this purpose because it excels in showing comparisons among discrete categories, such as different age ranges, on the X-axis. The Y-axis, on the other hand, is utilized to represent the count of employees within each age category. Its capacity to clearly depict the relationship between age categories and employee count enhances the interpretability for highlighting any trends or disparities in the age composition of the workforce, aiding in the assessment of age-related workforce characteristics.

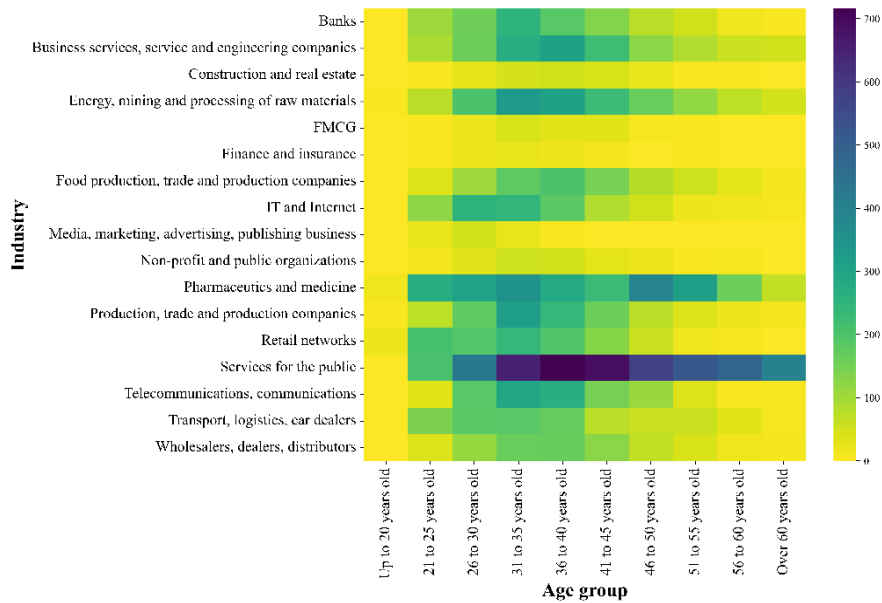

**Fig.4.** Heatmap of age ranges vs. industry sectors with workforce color-coded counts

A Heatmap, also known as a correlation map, is shown in **Fig.4** to examine the connections between various age groups to different industry sectors. With age ranges on the X-axis and industry sectors on the Y-axis, the color-coded representation, guided by a count-mark legend ranging from 0 to 700, allows for a detailed examination of the workforce distribution. The Heatmap is an ideal choice for this representation due to its ability to highlight patterns and variations across two categorical variables – age groups on the X-axis and various industry sectors on the Y-axis. This visualization method facilitates the quick identification of trends, revealing how age demographics vary across different sectors. For instance, the Heatmap shows that there are more employees in the Public Services sector in the age ranges of 31 to 35, 36 to 40, and 41 to 45, compared to FMCG, Finance & Insurance, which has lower employees count overall.

### 5.2    Work Satisfaction & Engagement

Understanding employee satisfaction and engagement within their current workplaces is a crucial aspect of our research. To assess this, we selected several pertinent questions from the survey (namely, Sr. No. {Questions} 12, 14, 15, 17, 18, 19, 30, 31, 34, and 37 from **Table 1**). Each of these questions targets specific aspects related to work satisfaction and engagement, including employees' interest in their work, their perception of their abilities, their satisfaction with their salary, their perceived value to the company, and their opinion on the company's appraisal methods [25].

While frequency plots for each question can yield important insights, we have chosen to employ violin plots for their ability to clearly delineate the distribution of responses across multiple categories. In this case, the responses included '*Agree*', '*Rather agree than disagree*', '*Disagree rather than agree*', '*Disagree*', and '*Difficult to answer*'.

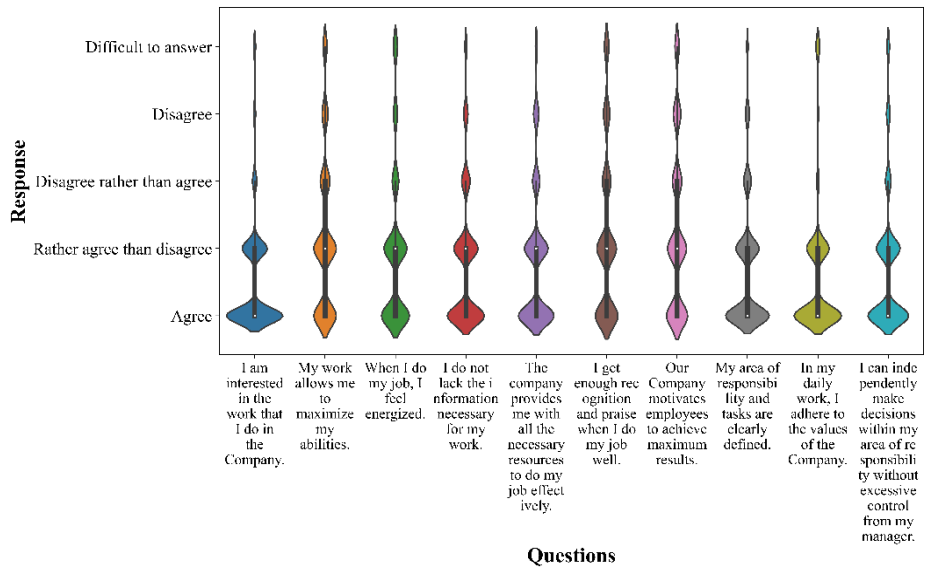

**Fig.5.** Violin plot showing response distribution of each question in the work satisfaction

This violin plot is shown in **Fig.5**, with each question tag on the X-axis and the corresponding responses on the Y-axis. The resulting visualization provides a clear picture of the response distribution for each question. Violin plots are a well-suited choice for this task as they excel in revealing the distribution of responses across various categories, making them particularly effective for survey data. The Violin plots offer a clear and intuitive representation, distinctly illustrating that the majority of responses cluster around '*Agree*' and '*Rather agree than disagree*', with a narrowing of the plots for the remaining response categories. This graph allows us to gauge general work satisfaction and engagement trends across the surveyed employees.

### 5.3    Work Environment Analysis

A comprehensive work environment analysis involves evaluating a variety of aspects [23]. We have selected several survey questions (specifically, Sr. No.{Que.} 13, 15, 29, 42, 45, 60, 63, and 65 from **Table 1**) that closely relate to this hypothesis. The selected questions revolve around factors such as organized work culture, provided benefits, opportunities for advancement, comfort at the workplace, and relationships with managers, and camaraderie with teammates, among others. Radar plot has its unique capability to effectively assess response density and patterns in work environment-related questions. The Radar plot excels in this context due to its ability to simultaneously represent multiple response categories ('Agree' to 'Disagree') in a single plot, each differentiated by a distinct color. This visualization choice offers a perfect view of response frequencies for specific questions, aiding in the precise evaluation of work environment factors by capturing density and distribution patterns. Hence, a radar plot presents the most compelling and comprehensive visualization for response analysis.

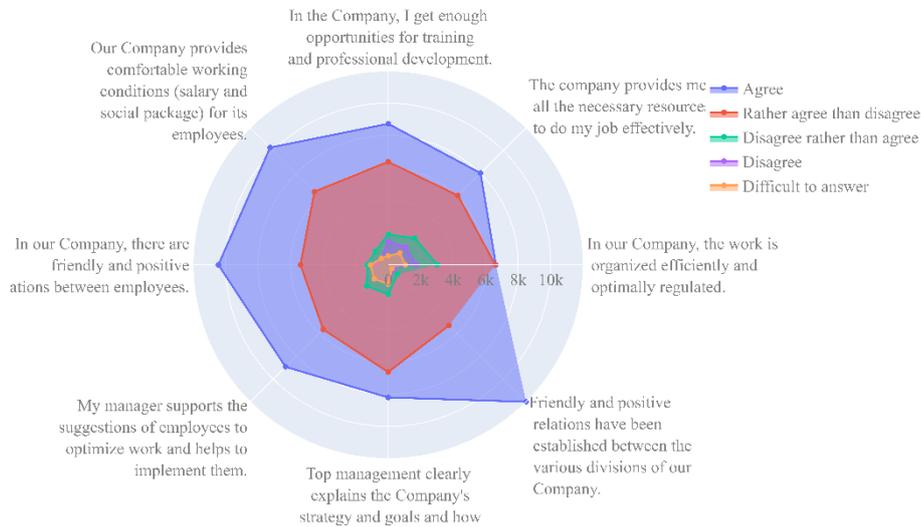

**Fig.6.** Radar plot illustrating frequency of responses to work environment related questions

In **Fig.6**, the spider graph within the radar plot illustrates the total count of responses, ranging from '*Agree*' to '*Disagree*', each represented by a distinct color. The perimeter of the radar plot displays the corresponding questions, while the legend accurately depicts the response categories. This graphical representation facilitates a clear understanding of the survey responses related to work environment factors.

### 5.4   Word cloud Analysis

A word cloud provides a striking visual representation of textual data, with words that appear more frequently in a source text given greater prominence. In **Fig.7**, we have created two distinct word clouds A & B, focusing on Questions 71 and 72 from the survey, which respectively address 'What makes our company a good employer?' and 'What could be improved in the work/processes in the company?'

In addition, to uncover common themes between the responses to these two questions, we utilized a Venn diagram approach, depicted in **Fig.7 (C).** This diagram presents overlapping categories between the two word clouds, helping identify key areas that can mutually enhance both the employee experience and company processes. Interestingly, the terms 'salary', 'team', 'training', 'staff', 'responsibility', 'work', 'time', 'organization', 'management', and 'company' emerge as common elements in both contexts. These words suggest focal points for both parties to work on and improve, enhancing the overall work environment and organizational processes.

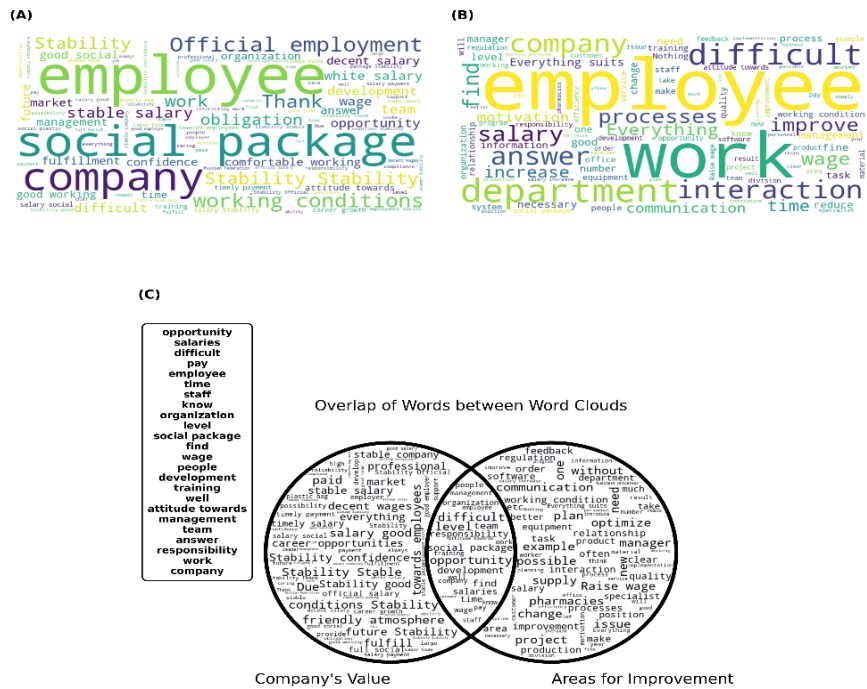

**Fig.7.** Word clouds for (A) factors contributing to employer attractiveness, (B) suggested areas for company's improvement, and (C) a Venn diagram identifying overlapping terminology

Word clouds are ideal for this purpose as they highlight the most frequently occurring terms, providing a quick and intuitive overview of the survey's textual responses. Their significance in employee surveys extends beyond this study, as they not only aid in current analysis but also open avenues for future research, particularly for NLP enthusiasts, by offering rich insights into employee sentiment and areas of concern.

### 5.5   Burnout Prediction

Burnout, a state of emotional, mental, and often physical exhaustion brought on by prolonged or repeated stress by any mean. High levels of burnout can also drive employees to consider leaving their current organization [10]. In our survey, Question 49 & 28 from **Table 1**, specifically probes the burnout levels among individual employees.

**Fig.8** breaks down key aspects of employee engagement in various sectors. It uses three components: Employee Satisfaction (shown in blue), Employee Burnout (in red), and Employee Attrition (in green). Employee Satisfaction is determined by counting "Agree" responses in Que_tag No 28, indicating the level of satisfaction. Employee Burnout is identified by counting "Agree" responses in Que_tag No 49, indicating an intention to leave within a year. Employee Attrition score is calculated using custom algorithm that link responses from both Questions, helping us accurately estimate potential burnout and attrition risk among employees. This figure provides a clear snapshot of these critical factors in employee engagement across different sectors.

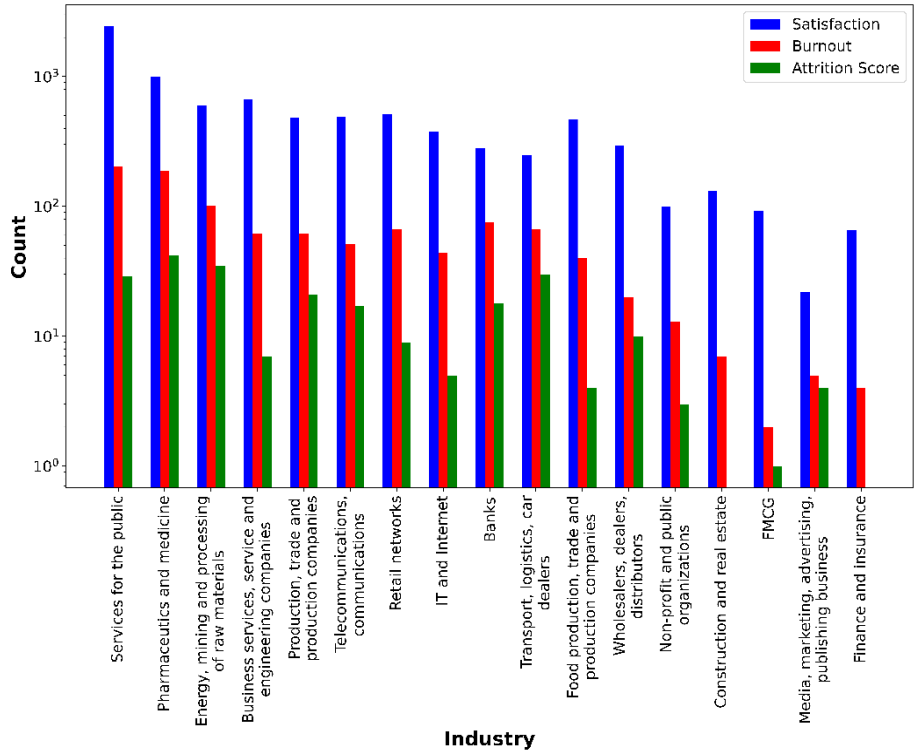

**Fig.8.** Industry-wise comparison of satisfaction, burnout, and employee attrition score

16      Bhimani et al.

Bar charts are ideal for this purpose as they distinctly represent the count of such employees in each industry, providing a clear visual of burnout likelihood with satisfaction. Also, underscoring their importance in quantifying and visualizing employee attrition score estimation, making them a precise choice for this aspect of our analysis.

To further substantiate our findings, we created a scatter plot illustrated in **Fig.9**, which considers all 60 questions from the survey. Scatter plot as a critical component of our analysis to investigate clusters and validate selected burnout and satisfaction features data in relation to survey responses. This plot includes a legend for responses and features the total count on the X-axis, with each question on the Y-axis. This comprehensive visualization offers detailed insights into each question and its corresponding hypothesis concerning the employee-organization relationship, thereby enriching our understanding of the burnout phenomenon and its possible ramifications in the workplace.

This visualization is pivotal in enriching our understanding of burnout in the workplace, offering a nuanced perspective on how survey responses align with our research objectives. The Scatter plot's significance lies in its ability to uncover patterns and relationships among variables, making it a valuable tool in this context.

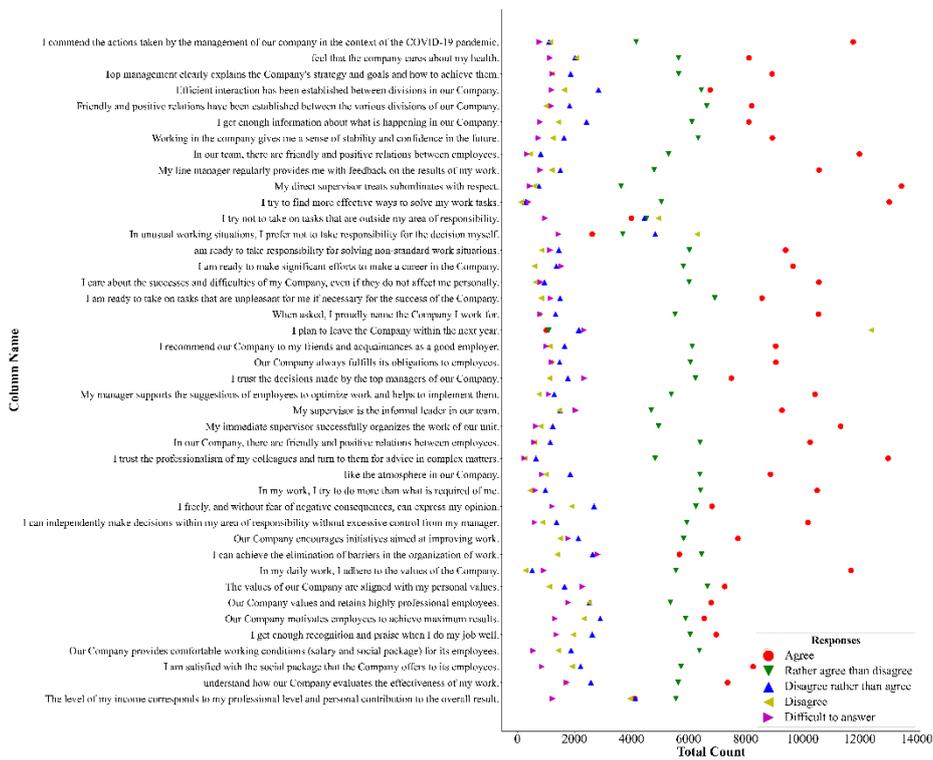

**Fig.9.** Scatter plot to find clusters and validate burnout with responses to survey question

## 6      Conclusion

This research stands as a testament to the profound implications of data science across various management sectors [5], particularly within the realm of employee-organization relationships. Through using this approaches one can enhance their findings. This broad-based approach is designed to resonate with a diverse array of readership. From researchers in sociology and management to firms seeking a detailed understanding of their workforce's satisfaction, our findings hold impact.

By leveraging a comprehensive dataset from a nationwide employee survey, we have delved deep into the multifaceted dynamics of the workplace, shedding light on critical aspects such as employee satisfaction, engagement levels, work environment conditions, demographic analysis, and the often overlooked issue of burnout prediction. Our approach in the realm of burnout research sets us apart. Rather than focusing solely on employee attrition scores, we have developed an algorithm that simultaneously assesses employee satisfaction and burnout levels, offering a holistic perspective that goes beyond traditional methods.

In summary, this research study serves as a bridge between the domains of data science and management, sociology, human resources, and administration. It encourages professionals in these sectors to embrace visualization-based approaches, fostering innovation and providing actionable insights that can drive positive change within organizations.

## 7      Discussion & Future Scope

*Current Limitations:* Our dataset, though extensive, primarily focuses on standard visualizations for employee survey and demographic data. However, we acknowledge that there is immense potential for further research in areas such as correlation, clustering, employee behavior, and burnout analysis. These avenues present opportunities for utilizing modern data science technologies and tools to inform organizational policies. Moreover, we recognize the value of developing a common data science pipeline that can be applied to various employee survey datasets, simplifying future analyses.

*Possible Future Directions:* The richness of our dataset indeed lends itself to various advanced techniques, including machine learning and deep learning algorithms. These can be harnessed for predicting burnout, assessing employee satisfaction, generating recruitment ideas, and identifying booming job market sectors. Additionally, text-based data offers an exciting frontier for Natural Language Processing (NLP) researchers to uncover behavioral insights. Furthermore, we intend to explore data visualization tools like Tableau and Power BI to create informative dashboards.

*Potential Challenges:* Given the vastness of data, it's crucial to maintain a keen awareness of data distribution and central tendencies to avoid introducing bias into analysis.

In our study, we have made concerted efforts to utilize a multitude of visualization techniques to comprehensively analyze the workplace culture and the interplay of employees within an organization. However, the exploration of this rich dataset and the attendant insights that it can provide does not end here. There exist multiple avenues to further this research. The canvas of possibilities ranges from discovering unique patterns and correlations in the data to apply more advanced methods of ML, DL, and NLP) models and making common toolkit for survey based analysis [13]. Such approaches can extract more nuanced insights and may allow for more precise predictions. Researchers can access, utilize, and download these resources at the following link: https://github.com/MrBhimani/Employee_Survey_Analysis

4